\documentclass[12pt]{amsart}
\usepackage{amsmath}

\usepackage{bbm,dsfont}

\newcommand{\va}{\mathbf{a}} %a
\newcommand{\vb}{\mathbf{b}} %b
\newcommand{\ve}{\mathbf{e}} %e
\newcommand{\vr}{\mathbf{r}} %r
\newcommand{\vn}{\mathbf{n}} %n
\newcommand{\vsigma}{\boldsymbol{\sigma}} %sigma
\newcommand{\cal}{\mathcal}

\newcommand{\eh}{{\cal E}\,({\cal{H})}}

\newcommand{\ran}{{\text{ran}\,}}
\newcommand{\lh}{{\cal L}({\cal H})}

\newcommand{\hi}{{\cal H}}
\newcommand{\lk}{\cal L(\cal K)}
\newcommand{\ki}{\cal K}

\newcommand{\ip}[2]{\left\langle\,#1\,|\,#2\,\right\rangle}

\newcommand{\kb}[2]{|#1\,\rangle\langle\,#2|}
\newcommand{\fii}{\varphi}

\newcommand{\tr}[1]{\mathrm{tr}\bigl[#1\bigr]}
\newtheorem{theorem}{Theorem}[section]
\newtheorem{lemma}[theorem]{Lemma}%[section]
\newtheorem{example}[theorem]{Example}%[section]
\newcommand{\R}{\mathbb R}
\newcommand{\br}{\mathcal B(\mathbb R)}

\newcommand{\N}{\mathbb N}

\newcommand{\sfq}{\mathsf{Q}}%position spectral measure
\newcommand{\sfp}{\mathsf{P}}%momentum spectral measure

\newcommand{\B}{\mathcal{B}}%generic observable
\newcommand{\F}{\mathcal{F}}%generic observable

\begin{document}
\title[Coexistence]{On the notion of coexistence in quantum mechanics}
\author{Paul Busch}
\address{Paul Busch,
Department of Mathematics, University of York, York, YO10 5DD, UK}
\email{pb516@york.ac.uk}
\author{Jukka Kiukas}
\address{Jukka Kiukas,
Institut f\"ur Theoretische Physik,
Universit\"at Hannover, 30167 Hannover,
Germany}
\email{Jukka.Kiukas@itp.uni-hannover.de}
\author{Pekka Lahti}
\address{Pekka Lahti,
Department of Physics and Astronomy, University of Turku, FIN-20014 Turku, Finland}
\email{pekka.lahti@utu.fi}

\begin{abstract}
The notion coexistence of quantum observables was introduced  to describe the possibility
of measuring two or more observables together. Here we survey the various different 
formalisations of this notion and their connections. We review examples illustrating the necessary 
degrees of unsharpness for two noncommuting observables to be jointly measurable (in one sense 
of the phrase). We demonstrate the possibility of measuring together (in another sense of the phrase) 
noncoexistent observables. This  leads us to a reconsideration of the connection between joint 
measurability and noncommutativity of observables and of the statistical and individual aspects of 
quantum measurements.

\

\end{abstract}

\maketitle

\section{Introduction}

The dual notions of states and observables 
are the basic  ingredients for formulating the probability structure of quantum mechanics.
If $\hi$ is the (complex separable) Hilbert space associated with the quantum system, then
the quantum mechanical (Born) probability formula is given by the  trace formula $p^E_\rho(X)=\tr{\rho E(X)}$;
here $\rho$ is the state  of the quantum system, a  positive trace one operator acting on $\hi$,
and $E:X\mapsto E(X)$ is the measured observable of the system, represented as
a semispectral measure on a $\sigma$-algebra $\mathcal A$ of  subsets of a set $\Omega$ with
positive, unit bounded operators $E(X)$ acting on $\hi$ as values. In this way any observable $E$ can be
identified with the map $\rho\mapsto p^E_\rho$, that is, with the totality of its associated
measurement outcome probability distributions.

The question of the possibility of measuring together (or jointly) two or more physical quantities 
lies at the heart of quantum mechanics. 
Insofar as the purpose of a measurement is to determine the probabilities for the various possible values of
the measured observable, this question amounts to asking the following:

\begin{quote}
\item[(Q)]
 Given any two observables $\rho\mapsto p^{E_1}_\rho$ and $\rho\mapsto p^{E_2}_\rho$,
with the value spaces $(\Omega_1,\mathcal A_1)$ and $(\Omega_2,\mathcal A_2)$, respectively, 
is there an observable $\rho\mapsto p^E_\rho$, with a value space $(\Omega,\mathcal A)$, from which $E_1$ and $E_2$ can be reconstructed in an operationally feasible way?
\end{quote}

There are three approaches  which have  been used extensively
to analyse the question (Q). The first one has its origin in the theory of sequential measurements,
the second refers  directly to joint measurements, whereas the third arises from the functional calculus of 
observables. Question (Q) can be rephrased accordingly in three ways.
\begin{itemize}
\item[1)] For which pairs of observables $E_1,E_2$ is the following statement true:
for any $\rho$, there is a probability bimeasure 
$$\mathcal A_1\times\mathcal A_2\ni(X,Y)\mapsto p_\rho(X,Y)\in[0,1]$$
such that $p^{E_1}_\rho$ and $p^{E_2}_\rho$ are its marginal measures, in the sense that
 $p_\rho(X,\Omega_2)= p^{E_1}_\rho(X)$ and  
$p_\rho(\Omega_1,Y)= p^{E_2}_\rho(Y)$
for any $X\in\mathcal A_1, Y\in\mathcal A_2$? If this is the case, $E_1$ and $E_2$ are said 
to have a {\em biobservable}, that is, 
there is a positive operator bimeasure 
$B:\mathcal A_1\times\mathcal A_2 \to\lh$ such that 
$E_1(\cdot) = B(\cdot,\Omega_2)$ and $E_2(\cdot)=B(\Omega_1,\cdot)$.
\item[2)]  For which pairs of observables $E_1,E_2$ is the following statement true:
for any $\rho$, there is a joint probability measure\footnote{$\mathcal A_1\otimes\mathcal A_2$ 
denotes the $\sigma$-algebra of subsets of $\Omega_1\times\Omega_2$
generated by the sets $X\times Y$, $X\in\mathcal A_1, Y\in\mathcal A_2$. } 
$$\mathcal A_1\otimes\mathcal A_2\ni Z\mapsto p_\rho(Z)\in[0,1]$$
such that 
 $p^{E_1}_\rho$ and $p^{E_2}_\rho$ are its marginal measures, that is,
 $p_\rho(X\times \Omega_2)= p^{E_1}_\rho(X)$ and  
$p_\rho(\Omega_1\times Y)= p^{E_2}_\rho(Y)$
for any $X\in\mathcal A_1, Y\in\mathcal A_2$? If this is the case, then $E_1$ and $E_2$
are said to have a {\em joint observable}, that is,  there  is an observable
$F:\mathcal A_1\otimes\mathcal A_2\to\lh$ such that  
$E_1(\cdot) = F(\cdot\times\Omega_2)$ and $E_2(\cdot)  = F(\Omega_1\times \cdot)$.
\item[3)]\label{i3} For which pairs of observables $E_1,E_2$ is the following statement true:
for any $\rho$, there is a probability measure $p_\rho$ defined on a $\sigma$-algebra $\mathcal A$
of a set $\Omega$ and measurable functions $f_1:\Omega\to\Omega_1$ and $f_2:\Omega\to\Omega_2$
such that $p_\rho(f_1^{-1}(X))=p^{E_1}_\rho(X)$ and $p_\rho(f_2^{-1}(Y))=p^{E_2}_\rho(Y)$
for any $X\in\mathcal A_1, Y\in\mathcal A_2$? If this is the case, $E_1$, $E_2$ are said
to be functions of $E$, in the sense that
$ E_1=E\circ f_1^{-1}$ and $E_2=E\circ f_2^{-1}$.
\end{itemize}
If $E_1$ and $E_2$ have a joint observable, then they also are functions of an observable, and if they are functions
of an observable, then they  have a biobservable. In general,  a biobservable is not  induced by a joint observable. 
However, if the measurable spaces 
involved are sufficiently regular, then such pathologies do not exist. 
Indeed, if the value spaces $(\Omega_1,\mathcal A_1), (\Omega_2,\mathcal A_2), (\Omega,\mathcal A)$ 
are Borel spaces, that is,
the sets 
are locally  compact metrizable and separable
topological spaces and the $\sigma$-algebras 
are the Borel $\sigma$-algebras,\footnote{Then also 
$\cal B(\Omega_1)\otimes \cal B(\Omega_2)=\cal B(\Omega_1\times\Omega_2)$} 
then  the three conditions are equivalent \cite{BCR84}, see also \cite{coex2,LY04}. 

\begin{example}\label{commutativity}\rm\small
As a first illustration, consider any two observables $E_1:\mathcal A_1\to\lh$ 
and $E_2:\mathcal A_2\to\lh$. If they  commute with each other, that is, 
$E_1(X)E_2(Y)=E_2(Y)E_1(X)$, for all $X\in\mathcal A_1, Y\in\mathcal A_2$, 
then the map $(X,Y)\mapsto E_1(X)E_2(Y)$ is a biobservable for $E_1$ and $E_2$.
If the value spaces are Borel spaces,  then $E_1$ and $E_2$ have a joint observable 
$F$ with the property $F(X\times Y)=E_1(X)E_2(Y)$.
If, in addition, one of the observables is projection valued, then $F$ is the unique joint 
observable of $E_1$ and $E_2$. This follows directly  from the fact that in such a case 
$E_1(X)E_2(Y)$ is the greatest lower bound of the effects $E_1(X)$
and $E_2(Y)$ \cite{MG1999}, for a slightly different argument, see, e.g. \cite{HRS09}.
\end{example}

Though important, the above  three reformulations of question (Q) do not exhaust its content. 
Below we shall describe yet another way of  phrasing and answering this question. 
Further, we will give examples of jointly measurable pairs of (generally noncommuting) 
observables and review some 
necessary and sufficient conditions for their joint measurability. This will enable us to identify
significant differences between the various notions of joint measurability considered here.

We start with a brief description of the notion of coexistence of observables, which 
has been  introduced  
as a seemingly obvious generalization of the idea of a joint observable 
for a pair of observables with finitely many values, and which encompasses the three notions
of joint measurability arising from the above formalisations of (Q).

\section{Coexistence}
Observables $E_1:\mathcal A_1\to\lh$ and $E_2:\mathcal A_2\to\lh$ are 
{\em coexistent}  if there is an observable $E:\mathcal A\to\lh$ such that
$$
\{E_1(X)\,|\, X\in\cal A_1\}\cup\{E_2(Y)\,|\, Y\in\cal A_2\}\subseteq\{E(Z)\,|\, Z\in\cal A\}.
$$
Such an observable $E$ will be called an {\em encompassing} observable for $E_1$ and $E_2$.
 Clearly,  
if $E_1:\cal A_1\to\lh$ and $E_2:\cal A_2\to \lh$ 
 have a  joint observable
or if they  are functions of  an observable,
then $E_1$ and $E_2$ are also coexistent. 
Moreover, if the value spaces involved are Borel spaces, then 
$E_1$ and $E_2$ are coexistent whenever they have  a biobservable.
In spite of many attempts \cite{coex1,coex2,coex3,DLPY05}
the question has remained open
whether the notion of coexistence 
is actually more general than these other three (essentially equivalent) notions of 
joint measurability.

Let $(\Omega,\cal A)$ and $(\Omega_1,\cal A_1)$ be any two measurable spaces. Then
for any observable $E:\cal A\to\lh$ and  measurable function $f:\Omega\to\Omega_1$, 
the range $\ran(E^f)$ of the image observable
$E^f:X\mapsto E^f(X)=E(f^{-1}(X))$ is contained in the range of $E$. The main problem is 
in the converse implication, that is:
if $E_1:\cal A_1\to\lh$ is an observable with the property  $\ran(E_1)\subseteq\ran(E)$, 
can one construct a function $f:\Omega\to\Omega_1$ such that 
$E_1=E^f$?\footnote{The condition $\ran(E_1)\subseteq\ran(E)$ need not imply that 
$E_1$ is a function of $E$; for an example, see \cite[Remark 1.1]{DLPY05}.} 
The classic results of Sikorski \cite{Sikorski} and Varadarajan \cite{V}, see also \cite{PP1991} 
and \cite{DLPY05}, show that such a construction is possible if the ranges are separable Boolean 
algebras. 
Example~\ref{Boolean} below is an application of this result. Yet the Boolean nature of the ranges 
of observables is not necessary for their functional calculus; some physically relevant examples 
have been studied in \cite{normi1}.
 Before recalling the Boolean case  we shall note another example where the above problem is 
 resolved, namely  the case where one of the observables is projection valued.

\begin{example}\label{PV}\rm\small
If $E_1$ and $E_2$ are coexistent, and if one of them  
is projection valued, then 
$E_1$ and $E_2$ commute with each other \cite[Th. 1.3.1, p. 91]{Ludwig83},
so that  the map $(X,Y)\mapsto E_1(X)E_2(Y)$ is a biobservable of $E_1$ and $E_2$.
If, in addition, the value spaces are Borel spaces, then they have  a joint observable $F$, 
which, by Example~\ref{commutativity} is necessarily   of the product form $F(X\times Y)=E_1(X)E_2(Y)$.
\end{example}

Let $\eh=\{A\in\lh\,|\, O\leq A\leq I\}$ be the set of effect operators. $\eh$ is equipped with the partial 
order $\leq$ (of selfadjoint operators) and the complementation map $A\mapsto A^\perp:=I-A$.
For an observable $E:\mathcal A\to\eh$, the range $\ran(E)=\{E(X)\,|\, X\in\cal A\}$  is not, in 
general,  a Boolean sub-$\sigma$-algebra of $\eh$, that is, the map $\cal A\ni X\mapsto E(X)\in\eh$ 
is not necessarily a $\sigma$-homomorphism, notwithstanding
the fact that $\cal A$ is a Boolean $\sigma$-algebra (of subsets of $\Omega$). 
It is an easy exercise to check that $\ran(E)$ is a Boolean subsystem of $\eh$ if an only if $E$ is 
projection valued.   
It may, however, happen that $E$ is a $\sigma$-homomorphism from $\cal A$ to 
$(\ran(E),\leq,\perp)$ without $E$ being projection valued. Indeed, for a given $E$
the system $(\ran(E),\leq,\perp)$ is Boolean if and only if $E$ is regular \cite{coex1,DLPY05}. 
We recall that $E$ is  
{\em regular} if there is no nontrivial effect operator $E(X)$ ($\ne O,I$) of $\ran(E)$ which would be 
either below $\frac 12I$ or above $\frac 12I$.

\begin{example}
\label{Boolean}\rm\small
If $E_1$ and $E_2$ are coexistent with an encompassing observable $E$ that is regular, 
then $\ran(E_1)$ and $\ran(E_2)$ are Boolean sub-$\sigma$-algebras of $\ran(E)$.
If the value spaces involved are complete separable metric spaces with the cardinality of $\R$, then 
$E_1$ and $E_2$  are functions of $E$. 
In particular, if the value spaces are real Borel spaces $(\R,\br)$, then the regularity of
an encompassing observable $E$ implies the existence of a biobservable $B$, a  joint observable 
$F$, and  Borel functions $f_1$ and $f_2$, such that 
$B(X,Y)=F(X\times Y)=E(f_1^{-1}(X)\cap f_2^{-1}(Y))$ for all $X,Y\in\br$, see e.g.  \cite{coex1,DLPY05}.
\end{example}

The problem with the notion of coexistence is that {\em in itself}, it does not entail a constructive
procedure for identifying an encompassing observable $E$ for $E_1,E_2$, nor for the embedding
of the ranges of the latter into the former. If it is given that observables $E_1$ and $E_2$  are coexistent 
with encompassing observable $E$, then all that is known is that there exists, for each $X\in\cal A_1$,
a set $Z_X\in\cal A$ such that $E_1(X)=E(Z_X)$, and similarly for $E_2$. On the basis of this information 
only, there seems to be no way to pick out the effect operators of $\ran(E_1)$ from those of $\ran(E)$, and 
similarly for $E_2$. Therefore, there seems to be no operational way to use the statistics $\rho\mapsto p^E_\rho$ to  
reconstruct the statistics of $E_1$ or $E_2$. 

By contrast, the notion of joint measurability does provide such a procedure and is, in addition, 
naturally adapted to the quantum mechanical modeling of measurement processes as we will recall next.

\section{Measurement theory}
According to the quantum theory of measurement, any observable (as a semispectral measure) 
$E:\cal A\to\lh$ admits a measurement dilation of the form
\begin{equation}\label{m1}
E(X)=V_\phi^*U^*I\otimes P(X)UV_\phi,
\end{equation}
where $U:\hi\otimes\cal K\to \hi\otimes\cal K$ is a unitary operator modelling the measurement 
coupling between the measured system
(with the Hilbert space $\hi$) and the apparatus (or the probe system, with the Hilbert space 
$\cal K$), $V_\phi$ is the embedding $\hi\to\hi\otimes\cal K$, $\fii\mapsto \fii\otimes\phi$, with 
$\phi$ being the initial probe (vector) state, and $P:\cal A\to\lk$ is the probe observable (which 
can be taken to be a spectral measure). We let $\cal M= (\ki,P,U,\phi)$ denote the measurement 
realization (\ref{m1}) of the observable $E$.

Let $(\Omega_1,\cal A_1)$ be any other measurable space, and let $f:\Omega\to\Omega_1$ 
be a measurable function, called a pointer function. The pointer function $f$ and the measurement 
$\cal M$ define another observable $E_1$, obtained as the image 
of $E$ under $f$,
\begin{equation}
E_1(X)=E(f^{-1}(X)), \quad X\in\cal A_1.
\end{equation}
Clearly, $\ran(E_1)\subseteq\ran(E)$, and, although $\cal M$ is not an $E_1$-measurement, the 
measurement $\cal M$ together with the pointer function $f$ constitutes a measurement of $E_1$. 
In particular, if any two observables $E_1$ and $E_2$ are functions of a third observable $E$, then 
any $E$-measurement $\cal M$ serves also as a measurement of both $E_1$ and $E_2$.

It may occur that one can use the measurement statistics  to construct the statistics 
of another observable without using such a functional calculus. We  describe next such a possibility.

\section{The method of moments}

We now review a possibility of determining the statistics of observable from the statistics of another
observable without the use of a functional calculus.

The method of moments refers to a case where from the moments of 
the actually measured statistics   
one is able to  infer the moments, 
and eventually the whole statistics
of another observable. 
Typically, such a situation arises 
when the actually performed measurement constitutes an unsharp measurement
of another obervables.

To describe this method in more detail,  let 
$E_{\fii,\psi}$ denote the complex measure $Y\mapsto E_{\fii,\psi}(Y)=
\ip{\fii}{E(Y)\psi}$ defined by  an observable   $E:{\cal B}(\R)\to L(\hi)$ and the vectors $\fii,\psi\in\hi$. 
In particular, if $\fii\in\hi$ is a unit vector, then $E_{\fii,\fii}=p^E_\rho$,
with $\rho=\kb{\fii}{\fii}$.
We recall that the $k$-th moment operator $E[k]$ of $E$ is  the weakly defined operator
$E[k]=\int_\R x^k\,dE(x)$, with the domain $D(E[k])$ consisting of those vectors $\psi\in\hi$
for which the integral $\int x\,dE_{\fii,\psi}(x)$ exists for  all $\fii\in\hi$. In particular, if
the integrals $\int_\R x^k\, dp^E_\rho(x)$ exist, they define the moments of the measurement outcome 
statistics $p^E_\rho$.

Let  $\mu:{\cal B}(\R)\to [0,1]$ be a  probability measure, and let
$\mu*E$ denote the convolution of  $\mu$ and $E$. It is the
observable  $X\mapsto (\mu*E)(X)$  defined by 
$\langle \fii|(\mu*E)(X)\psi\rangle = \mu*E_{\fii,\psi}(X)$, $\fii,\psi\in \hi$,
where $\mu*E_{\fii,\psi}$ is the convolution of $\mu$ with
the complex measure $E_{\fii,\psi}(Y)$, that is, 
\begin{equation}
\mu*E_{\fii,\psi}(X) = \int_\R \mu(X-y)\, dE_{\fii,\psi}(y).
\end{equation}

We note that
$\ran(E)$ is contained in $\ran(\mu*E)$ only if $\mu$ is a point measure. 
However, it may 
happen that one can reconstruct (the moments of) $E$ from (the moments of) $\mu*E$ in such a way that the full statistics become uniquely determined. 
Indeed, the moment operators of $\mu*E$ and $E$ are related with each other as follows \cite{KLY08}.

\begin{lemma} Let $E:{\cal B}(\R)\to L(\hi)$ be a semispectral
measure, and $\mu:{\cal B}(\R)\to [0,1]$ a probability measure.   
If $\mu[k]$ exists, then ${\cal D}(E[k])\subset {\cal D}((\mu*E)[k])$, and
\begin{equation}\label{mm1}
(\mu*E)[k] \supset \sum_{n=0}^k \binom{k}{n} \mu[k-n]E[n].
\end{equation}
\end{lemma}

Assume now that all the moments $\mu[k]$ of  the blurring probability measure $\mu$ are finite and
in addition that $\emptyset\not= D\subset\cap_{k=0}^\infty  {\cal D}(E[k])$. 
Then for any state $\rho=\kb{\fii}{\fii}$, $\fii\in D$, 
\begin{equation}\label{mm2}
p^{\mu*E}_\rho[k] = \sum_{n=0}^k \binom{k}{n} \mu[k-n]p^E_\rho[n].
\end{equation}
from which a recursion formula for the moments $p^E_\rho[k]$ is obtained:
\begin{equation}\label{mm3}
p^{E}_\rho[k] = p^{\mu*E}_\rho[k] - \sum_{n=0}^{k-1} \binom{k}{n} \mu[k-n]p^E_\rho[n].
\end{equation}
Assume, further, that the moments $p^{E}_\rho[k]$ fulfill the operationally verifiable condition
\begin{equation}\label{mm4}
|p^{E}_\rho[k]| \leq CR^k k!, k=1,2,\ldots.
\end{equation}
This implies that $\int e^{a|x|}\,dp^E_\rho<\infty$, 
whenever $0<a<1/2R$ (see, e.g. \cite[Proposition 1.5]{Simon}),
showing that the probability measure $p^E_\rho$
is exponentially bounded,
a condition which  assures that the moment sequence $(p^{E}_\rho[k])_{k=0}^\infty$
uniquely determines the probability measure $p^{E}_\rho$, see, e.g., 
\cite[p. 406, Theorem 30.1]{BillingsleyII}. If the set $D$ above is a dense subspace, 
then the probability measures $p^{E}_\rho$, $\rho=\kb\fii\fii,\fii\in D$,
determine, by the polarization identity, the observable $E$. Note that if $E$ is a spectral 
measure, then there always exists such a dense subset $D$ so that it only remains to check 
that the convolving measure $\mu$ has finite moments
and that the condition (\ref{mm4}) is satisfied for a sufficiently large set of states $\rho$.

We conclude that under the operational conditions specified above
one can reconstruct the moments of $p^{E}_\rho$ and then uniquely determine the corresponding statistics, from   the actually measured distribution $p^{\mu*E}_\rho$ 
using the method of moments
even though the range of $E$ may not be contained in the range of  ${\mu*E}$.
One may call this an {\em indirect} $E$-measurement.

\section{Examples: measuring together sharp noncommuting observables}

\subsection{Indirect measurement of sharp position and momentum}\label{subsec:indir-qp}
Let $\sfq$ and $\sfp$ denote the spectral measures of the position and momentum operators $Q$ 
and $P$, acting in $L^2(\R)$. The convolutions $\mu*\sfq$ and $\nu*\sfp$ of $\sfq$ and $\sfp$ 
with probability measures  $\mu,\nu:\br\to[0,1]$ are unsharp position and momentum observables, 
respectively. The standard (von Neumann) model of a position measurement  constitutes a realization of
$\mu*\sfq$ where $\mu$ is an absolutely continuous probability measure depending on the preparation
of the measurement probe and the coupling strength between probe and particle (for details, see \cite{OQP}). 

Consider a measurement of $\mu*\sfq$.
Assuming that all the moments of  $\mu$ are finite, for instance, in the standard model the initial 
probe state is a (compactly supported) slit-state or a Gaussian state, and choosing $D$ to be, for 
instance, the linear span of the (normalized) Hermite functions, 
then the moments $p^{\sfq}_\rho[k], k\in\N$, can be obtained  recursively by  (\ref{mm3}) from
 the actually measured statistics $p^{\mu*\sfq}_\rho$, $\rho=\kb\fii\fii, \fii\in D$, and they fulfill the 
condition (\ref{mm4}).
Therefore, the numbers $p^{\sfq}_\rho[k], k\in\N$, 
determine the distribution $p^{\sfq}_\rho$. Due to the density of $D$, the actual measurement
$\rho\mapsto p^{\mu*\sfq}_\rho$ thus determines the whole observable $\rho\mapsto p^\sfq_\rho$,
that is, the (sharp) position observable $\sfq$ is measured indirectly by a measurement of an 
unsharp position $\mu*\sfq$, whenever all moments of the blurring measure $\mu$ are finite.

Similarly, an unsharp momentum measurement can serve as an indirect measurement of the sharp 
momentum.

Sharp position and momentum observables $\sfq$ and $\sfp$ are (strongly) noncommutative.
Therefore, they are  noncoexistent, they do not have a biobservable or joint observable, nor are 
they functions of a third observable.  Nevertheless they can be 
measured together indirectly, that is, there are measurements $\mathcal M$ which allows one to 
determine, from the actual statistics $\rho\mapsto p^E_\rho$, both  the position
statistics $\rho\mapsto p^\sfq_\rho$ and the momentum statistics $\rho\mapsto p^\sfp_\rho$.

The Weyl operators representing phase space translations by a displacement 
$(q,p)\in\R^2$ are defined as
$W_{qp}=e^{i\frac 12 qp}e^{-iqP}e^{ipQ}$. 
It is well known that any covariant phase space observable  $G^T$ is generated by a
positive operator $T$ of trace one (acting in $L^2(\R)$) such that for $Z\in\mathcal B(\R^2)$,
\begin{equation}
G^T(Z)=\frac 1{2\pi}\int_ZW_{qp}TW_{qp}^*\,dqdp\,.
\end{equation}
The Cartesian marginal observables of $G^T$ are the unsharp position and momentum 
observables $\mu*\sfq$ and $\nu*\sfp$, with  $\mu$ and $\nu$ defined by the Fourier related 
densities $f(q)=\sum_it_i|\eta_i(-q)|^2$ and $g(p)=\sum_it_i|\hat\eta_i(-p)|^2$,
where $T=\sum_i t_i\kb{\eta_i}{\eta_i}$ is the spectral decomposition of the generating operator $T$.
Choosing the generating operator $T$ such that all the moments $\mu[k]$ and $\nu[k]$, $k\in\N$, 
are finite, and using $D$ as given above, we conlude that the marginal measurement statistics 
$\rho\mapsto p^{\mu*\sfq}_\rho$ and  $\rho\mapsto p^{\nu*\sfp}_\rho$, $\rho=\kb\fii\fii, \fii\in D$, 
collected under a single measurement scheme, suffice to determine both  the position statistics 
$\rho\mapsto p^\sfq_\rho$ and the momentum statistics $\rho\mapsto p^\sfp_\rho$.

The Arthurs-Kelly model, or a sequential standard position measurement followed by a sharp 
momentum measurement,  or the eight-port homodyne detection scheme provide examples of 
measurement realizations of such a joint determination of the position and momentum statistics. 
For a more detailed discussion of these models, see, for example,
\cite{Busch1982,Raymer,TSJ,OQP,JukkaVII, BKL08,KLS09}.

\subsection{Indirect measurement of spin-$\mathbf{\frac 12}$ components}

For observables with discrete or even finite sets of outcomes it becomes particularly simple
to consider the question of indirect measurements. Let $P_1,P_2,\dots, P_n$ be a complete
family of mutually orthogonal projections such that $\sum_kP_k=I$, and let $(\lambda_{jk}$
be a stochastic $m\times n$ matrix, that is, $\lambda_{jk}\ge 0$, $\sum_k\lambda_k=1$. Then
the operators $E_j=\sum_k\lambda_{jk}P_k$ are positive and satisfy $\sum_jE_{j}=I$, that is,
they constitute an observable which is a smeared version of the sharp observable defined by the $P_k$.
If the matrix $(\lambda_{jk})$ is square and invertible, it follows that the $P_k$ can be expressed as
linear combinations of the $E_j$, so that $P_k=\sum_j \mu_{kj}E_j$. It follows that  $\tr{\rho P_k}=\tr{\rho' P_k}$
for all $k$ if and only if $\tr{\rho E_j}=\tr{\rho' E_j}$; in other words, the observables given by 
$\{P_k:k=1,2,\ldots,n\}$ and 
$\{E_j:j=1,2,\ldots,n\}$ are equally good at separating distinct states, they are informationally 
equivalent. 

As an example, we consider the joint determination of the statistics of noncommuting spin components
of a spin-$\frac 12$ system.
Using the bijective correspondence between $M(\mathbb C^2)$ and $\mathbb C^4$ mediated by the 
Pauli basis $I,\sigma_1,\sigma_2,\sigma_3$, we recall that any operator ($2\times 2$ matrix) can 
be written as $A=a_0I+\va\cdot\vsigma$. A state $\rho$ is given by $\rho=\frac 12(I+\vr\cdot\vsigma)$, 
with $\vr\in\mathbb R^3$, $|\vr|\le1$. The following four-outcome observable $G$ is an example of a 
joint observable for smeared 
versions of the sharp  observables $\sigma_1$ and $\sigma_2$:
\begin{equation}
\{+,-\}\times\{+,-\}\ni (j,k)\mapsto G_{jk}=\tfrac 14(I+\vn_{jk}\cdot\vsigma)\,,
\end{equation}
where $\vn_{+,\pm}=(\ve_1\pm\ve_2)/\sqrt{2}$, $\vn_{-,\pm}=(-\ve_1\pm\ve_2)/\sqrt{2}$. The two obvious marginal
observables are given by the following pairs of effects:
\begin{eqnarray}
E^{(1)}_\pm&=&G_{\pm,+}+G_{\pm,-}=\tfrac 12( I\pm\tfrac{\sqrt 2}{2}\sigma_1)\\
&=&\tfrac 12(1\pm\tfrac{\sqrt 2}2)P^{(1)}_++
\tfrac 12(1\mp\tfrac{\sqrt 2}2)P^{(1)}_-\,,\nonumber
\end{eqnarray}
\begin{eqnarray}
E^{(2)}_\pm&=&G_{+,\pm}+G_{-,\pm}=\tfrac 12( I\pm\tfrac{\sqrt 2}{2})\sigma_2)\\
&=&\tfrac 12(1\pm\tfrac{\sqrt 2}2)P^{(2)}_++
\tfrac 12(1\mp\tfrac{\sqrt 2}2)P^{(2)}_-\,.\nonumber
\end{eqnarray}
Here the $P^{(k)}_\pm:=\frac 12(I\pm\sigma_k)$ are the spectral projections of $\sigma_k$.
It is obvious that the transformations $\{P^{(k)}_\pm\}\to \{E^{(k)}_\pm\}$ are invertible. In fact, we have:
\begin{equation}
P^{(k)}_\pm=\pm\tfrac 12(\sqrt 2\pm1)E^{(k)}_+\mp\tfrac 12(\sqrt 2\mp1)E^{(k)}_-\,.
\end{equation}
Thus the statistics $p^{\sigma_k}_\rho$, $k=1,2$ can be reconstructed from the statistics
of $E^{(k)}$ via measuring the joint observable $G$.

Realistic models of the measurement of a observable of the form $\{G_{jk}\}$ were first presented in 
\cite{Busch1987}. A systematic study of  the reconstruction of sharp spin-$\frac 12$ observables
from such non-ideal or approximate joint measurements can be found in \cite{Ma-deMu90}.

\section{The method of state reconstruction}

The state $\rho$ of a quantum particle  is not determined by its position and momentum distributions 
$p^\sfq_\rho$ and $p^\sfp_\rho$. This is a  well-known but  important  nonclassical feature of the 
quantum theory, a feature  called   {\em surplus information} by C.F. von Weizs\"acker \cite{CFvW85}.

If an observable $E$ is informationally complete, that is, the map 
$\rho\mapsto p^E_\rho$ separates states, and if one can determine an algorithm for reconstructing 
the state $\rho$ from the statistics $p^E_\rho$, then one can obtain the measurement statistics of any
observable, in particular the statistics $p^\sfq_\rho$ and $p^\sfp_\rho$. Feasible state reconstruction algorithms are known for quadrature distributions\footnote{Note that these can easily be bunched together to form a single informationally complete observable.}, and simple phase space observables (see \cite{DAriano}, and also \cite{KLP2008}). Hence, it is clear that the question (Q) can have a positive answer without the observables being coexistent.

The reconstruction scheme of subsection \ref{subsec:indir-qp} shows, however, that
there are single measurement schemes which allow one to reconstruct, in an operational way, the moments of the distributions  $p^\sfq_\rho$ and $p^\sfp_\rho$ without  the need to reconstruct first the state $\rho$; in fact, in the example of a phase space observable $G^T$, it is not necessary to require that $G^T$ is informationally complete.\footnote{The informational completeness of $G^T$ is known to be equivalent to
the assumption that  
 $\tr{W_{qp}T}\ne 0$ for almost all $(q,p)\in\R^2$ \cite{Ali,JW}.}
For a suitable subclass of states, the moments contain the same information as the distributions themselves, so one may speak of a measurement. (Obviously, a serious disadvantage of this method compared to the state reconstruction is that the distributions cannot be algorithmically reconstructed from the moments.)

In the case of a spin-$\frac 12$ system, it is possible to obtain simultaneous reconstructions of the spin 
component observables $\sigma_{\vn}$ along {\em all} directions (specified by the unit vectors $\vn$)
from a single observable $M$ as follows. Let $(S^2,\cal B(S^2))$ denote the unit 2-sphere in $\mathbb R^3$,  
equipped with the standard Borel $\sigma$-algebra, and let $d\Omega(\vn)$ be the uniform surface 
measure normalised as $\Omega(S^2)=4\pi$. The following is a normalized positive operator measure:
\begin{equation}
\cal B(S^2)\ni Z\mapsto M(Z):=\frac 1{2\pi}\int_Z \tfrac12(I+\vn\cdot\vsigma)\,d\Omega(\vn)\,.
\end{equation}
It is obvious that one can define, for any direction, a 2-valued marginal observable by choosing a partition
of $S^2$ in the form of two complementary hemispheres with poles along the given direction. For simplicity,
we consider the direction along the $z$-axis of some Cartesian coordinate system, and denote the partition
as $Z_\pm$. Then we obtain:
\begin{equation}
M(Z_\pm)\equiv M_\pm=\tfrac 12(I\pm \tfrac 12\sigma_3)\,.
\end{equation}
As before, the statistics of $\sigma_3$ can be reconstructed from this marginal observable.

\section{Joint measurability of noncommuting observables}

The examples of the preceding sections illustrate the following well-known fundamental fact. Noncommuting
sharp observables do not admit any joint observable; but there are smeared versions of such noncommuting
observables that do possess joint observables. The natural question about the necessary amount of
unsharpness required to ensure joint measurability has not yet been answered in full generality, but important
insights have been gained in special cases.

In the case of position and momentum it is known \cite{Werner2004,CaHeTo2004} that observables %\marginpar{\bf Refs ok?}
$\mu*\sfq$ and $\nu*\sfp$ are jointly measurable if and only if the smearing measures $\mu,\nu$ have 
Fourier-related densities, in the sense described in subsection \ref{subsec:indir-qp}. In this case the variances of
these measures, which represent measures of the inaccuracies of the position and momentum determination, 
satisfy the Heisenberg uncertainty relation, 
\begin{equation}
{\rm Var}(\mu)\,{\rm Var}(\nu)\ge\frac\hbar 2\,,
\end{equation}
which thus is seen to constitute a necessary condition of joint 
measurability.

In the case of  a spin-$\frac 12$ system or, more generally, a qubit represented by a 2-dimensional Hilbert 
space $\mathbb C^2$, the question of necessary and sufficient conditions for the coexistence (equivalently, 
joint measurability) of a pair of effects has recently been completely answered \cite{BuSchm2009,StReHe2008,Yu-etal2008}. 
Effects 
$A=a_0I+\va\cdot\vsigma$ and $B=b_0I+\vb\cdot\vsigma$ are coexistent
if and only if a certain  inequality holds which can be cast in the following form \cite{BuSchm2009}:
\begin{equation}\label{eqn:coex-ineq}
\tfrac 12[\F(2-\B)+\B(2-\F)]+(xy-4\va\cdot\vb)^2\ge 1.
\end{equation}
Here the following abbreviations are used:
\begin{eqnarray}
\F&:=&\varphi(A)^2+\varphi(B)^2;\\
 \B&:=&\beta(A)^2+\beta(B)^2;\\
x&:=&\varphi(A)\beta(A)=2a_0-1;\label{eqn:x}\\
y&:=&\varphi(B)\beta(B)=2b_0-1;\\
\varphi(A)&:=&\sqrt{a_0^2-|\va|^2}+\sqrt{(1-a_0)^2-|\va|^2};\label{eqn:fa2}\\
\beta(A)&:=&\sqrt{a_0^2-|\va|^2}-\sqrt{(1-a_0)^2-|\va|^2}.\label{eqn:ba2}
\end{eqnarray}
$\varphi(B)$ and $\beta(A)$ are defined similarly. The quantity $\varphi(A)$ is a measure of the degree of
unsharpness of the effect $A$, and $\beta(A)$ and $x$ are measures of the  bias of
$A$. An effect $A$ (and its complement $A'=I-A$) is unbiased if the mid-point of its spectrum is $\frac 12$.
(A more detailed investigation of these properties and measures can be found in \cite{Busch2009}.) The
degree of noncommutativity is represented by the deviation of the term $|\va\cdot\vb|$ from $\|\va\|\,\|\vb\|$. This 
inequality represents a rather complicated trade-off between the unsharpness, bias and noncommutativity degrees
of the two effects $A,B$, which must hold if they are to be jointly measurable.

In the special case of unbiased effects ($a_0=b_0=\frac 12$), the above coexistence inequality assumes the simple 
form
\begin{equation}
16 |\va\times\vb|^2\le (1-4|\va|^2)(1-4|\vb|^2)
\end{equation}
Considering that $A=\frac12I+\va\cdot\vsigma$ is a projection if and only if $|\va|=\frac 12$, we see that $1-4|\va|^2$
is a measure of the unsharpness of $A$. Hence the product of the degrees of unsharpness of $A,B$ is bounded
below by the square of the vector product of $\va$ and $\vb$, which is proportional to the commutator of $A$ and $B$.

\section{Discussion}

We have reviewed three notions of joint measurability (based on biobservables, joint observables and
functional calculus, respectively), the notion of coexistence,  and a concept of indirect measurement,
and the known logical relations between these notions.

The three notions of joint measurability provide operationally feasible ways of 
reconstructing the statistics of two observables  $E_1,E_2$ from their joint observable. We went on to 
show that such operational reconstruction can even be achieved in cases where $E_1, E_2$ are 
{\em not} jointly measurable. We found important instances where from the statistics $p^E_\rho$ of 
a given  observable $E$ one may, without the need of full state reconstruction, infer the statistics of another observable $p^{E_1}_\rho$ even though 
$\ran(E_1)$ is not contained in $\ran(E)$, and, {\em a fortiori}, there is no functional connection between 
$E_1$ and $E$. For example, the statistics of sharp position and momentum can be recovered 
from the statistics of a single phase space observable for a dense set of states. Hence, also in this case, the question (Q) has a positive answer without the observables involved being coexistent. ``Measuring together"
two observable that are not jointly measurable thus amounts to a common indirect measurement of the
two observables, typically based on a measurement of a bi- or joint observable of  unsharp versions or approximations
of them.

For two observables that are coexistent or jointly measurable, there is an event, associated with an effect $E(Z_{X,Y})$
from the joint or encompassing observable $E$, that represents the joint occurrence of two effects $E_1(X)$ and
$E_2(Y)$. Such joint events do not exist for non-coexistent observables. This ``deficiency" cannot be removed
through a common indirect measurement. In this sense ``measuring together" two observables in an indirect measuremet
is a weaker notion than ``measuring jointly". A more quantitative description of the idea of measuring two  noncommuting 
sharp (hence non-coexistent) observables together ``indirectly" has been obtained in investigations of recent years
into a precise notion of {\em approximate joint measurement}. The examples given above, of joint measurements 
of unsharp versions of such sharp observables,  can be considered as approximate joint measurements. The
quality of the approximation is quantified by measures of inaccuracy, given by the distance between each one
of the sharp observables from one of the marginals of the joint observable. Measurement uncertainty relations for
such approximate joint measurements have been obtained, for example, in \cite{Werner2004,BuHeLa2007,MiIm2008}.

The existence of ``joint events" for effects $E_1(X),E_2(Y)$ of two jointly measurable observables represented 
by  effects $F(X\times Y)$ from a joint observable $F$ gives rise to an interpretation in terms of joint unsharp
values that can be prepared by a suitable choice of measurement of $F$. An example of a weakly preparing
measurement operation is given by the generalised L\"uders operation associated with an effect $A$, defined 
as the map \cite{Busch1987}
\begin{equation}
\rho\mapsto A^{1/2}\rho A^{1/2}\,.
\end{equation}

We also noted that while joint measurability implies coexistence, it is not known whether there are coexistent
observables that are not jointly measurable. Here we face the following two possibilities: It may be the case that 
coexistence is no more 
general than (one of) the three other notions of joint measurability; then the notion of coexistence 
adds no new possibilities. Alternatively, coexistence may turn out to be more general logically.
There would thus be pairs of observables that are coexistent although they are not jointly 
measurable; but there seems to be no operational way of obtaining the probability distributions of 
the two observables in question from the encompassing observable. We conclude that the notion 
of coexistence has no added value over and above the three other notions of joint measurability. 
As far as the operational possibilities of joint measurements are concerned, one can safely
use the term {\em coexistence} as a convenient synonym for {\em joint measurability}.

\end{document}